\documentclass[aps,pra,showpacs,twoside,twocolumn,10pt,longbibliography]{revtex4-2}
\usepackage[colorlinks=true, citecolor=red, urlcolor=blue ]{hyperref}
\usepackage{times,epsfig,amssymb,amsfonts,amsmath,bm,subfigure,mathtools,amsthm,braket,soul,enumitem,color, caption} 
\usepackage[normalem]{ulem}
\usepackage{graphicx}
\usepackage{xcolor}
\usepackage{subfloat}
\usepackage{float}
\usepackage{braket}
\usepackage{tabularx}
\usepackage{lipsum}
\usepackage{booktabs}
\usepackage{multirow}
\usepackage{natbib}
\newcolumntype{C}[1]{>{\centering\arraybackslash}p{#1}}

\newcommand{\stkout}[1]{\ifmmode\text{\sout{\ensuremath{#1}}}\else\sout{#1}\fi}
\definecolor{magenta}{rgb}{1.0, 0.0, 0.56}

\newcolumntype{M}[1]{>{\centering\arraybackslash}m{#1}}
\newcolumntype{N}{@{}m{0pt}@{}}

\begin{document}
\title{The role of data embedding in equivariant quantum convolutional neural networks}
\author{Sreetama Das$^{1, 2, 3}$, Stefano Martina$^{2, 3}$, Filippo Caruso$^{1,2,3}$}
\affiliation{$^1$Istituto Nazionale di Ottica del Consiglio Nazionale delle Ricerche (CNR-INO), I-50019 Sesto Fiorentino, Italy}
\affiliation{$^2$Department of Physics and Astronomy, University of Florence, Via Sansone 1, Sesto Fiorentino, I-50019, Italy}
\affiliation{$^3$European Laboratory for Non-Linear Spectroscopy (LENS), University of Florence, Via Nello Carrara 1, Sesto Fiorentino, I-50019, Italy}

\begin{abstract}
Geometric deep learning refers to the scenario in which the symmetries of a dataset are used to constrain the parameter space of a neural network and thus, improve their trainability and generalization. Recently this idea has been incorporated into the field of quantum machine learning, which has given rise to equivariant quantum neural networks (EQNNs). In this work, we investigate the role of classical-to-quantum embedding on the performance of equivariant quantum convolutional neural networks (EQCNNs) for the classification of images. We discuss the connection between the data embedding method and the resulting representation of a symmetry group and analyze how changing representation affects the expressibility of an EQCNN. We numerically compare the classification accuracy of EQCNNs with three different basis-permuted amplitude embeddings to the one obtained from a non-equivariant quantum convolutional neural network (QCNN). Our results show a clear dependence of classification accuracy on the underlying embedding, especially for initial training iterations. The improvement in classification accuracy of EQCNN over non-equivariant QCNN may be present or absent depending on the particular embedding and dataset used. It is expected that the results of this work can be useful to the community for a better understanding of the importance of data embedding choice in the context of geometric quantum machine learning.
\end{abstract}

\maketitle

\section{Introduction}

Quantum computing holds the promise to surpass classical supercomputers in achieving polynomial and exponential speed-up when performing certain tasks~\cite{shor_1997, grover_1996}. The recent breakthrough~\cite{google_supremacy} of realizing such speed-up with state-of-the-art noisy intermediate-scale quantum (NISQ) devices~\cite{Preskill2018nisq} has drawn widespread attention and research interest to this field. One of the most active research lines in recent times is to connect quantum computation with classical machine learning (ML). In this context there exist two approaches, the first one is to study how classical ML tools can facilitate quantum information processing tasks~\cite{preskill_science_2022, martina2022learning,martina2023machine,martina2023deep,canonici2023machine}. The other is the more frequently studied direction, i.e. to use a quantum system itself to build machine learning models~\cite{dalla2022quantum,das2023quantum} and in particular quantum neural networks (QNNs)~\cite{qml_schuld_2015, qml_lloyd_2017, Benedetti_2019, Perdomo-Ortiz_2018, schuld_2022_prx,parigi2023quantumnoisedriven,zhang2023generative,cacioppo2023quantum}. The central component in a prototypical QNN is a quantum circuit with single and multiple qubit gates with trainable parameters. The latter is called parametric quantum circuit (PQC) or variational quantum circuit (VQC)~\cite{peruzzo_2014, McClean_2016, Romero_2017, mitarai_2018, schuld_2020}. PQC-based QNNs have been used to design quantum analogs of well-known classical ML networks, e.g., quantum autoencoder~\cite{Romero_2017}, quantum convolutional neural networks (QCNNs)~\cite{cong_2019, Li_2020, quanvolution_2020}, quantum generative adversarial networks (QGANs)~\cite{Braccia_2021, braccia_2022, rudolph2022generation, boyle2023hybrid, tsang2023hybrid, ZHOU2023116891}, quantum generative diffusion models~\cite{parigi2023quantumnoisedriven,zhang2023generative,cacioppo2023quantum}, etc. It is important to note here that the cost function obtained from the above networks is optimized using a classical optimization routine. In this sense, the quantum machine learning networks are \textit{hybrid quantum-classical networks}. Despite the potential speed-up and signatures of success of quantum machine learning over classical ML models \cite{rebentrost_svm_2014, Liu_2021, abbas_2021, huang_science_2022, caro_2022}, there exist several limitations for QNNs, the \textit{barren plateau} problem being the principal among them. The latter means that for sufficiently deep quantum neural networks the gradient of the cost function vanishes exponentially with the size of the PQC~\cite{mcclean_barren_2018, holmes_2022, cerezo_2021, sharma_2021}. On one hand, an arbitrary PQC ideally should have high expressibility \cite{sim_2019} to ensure that the solution of the optimization problem is close enough to the actual solution. On the other hand, PQCs with higher expressibility are more prone to exhibiting barren plateau~\cite{mcclean_barren_2018, holmes_2022}. It is therefore a crucial task to mitigate barren plateau for a practical application of quantum machine learning algorithms.

One way to improve the trainability and generalization in machine learning algorithms is to introduce \textit{inductive bias} in the network, i.e. to use some prior known information about the dataset to build a problem-specific model and constrain the optimization space of the network. Particularly, \textit{geometric machine learning} refers to a scheme in which the known symmetries of the dataset are used to construct a network which respects those symmetries~\cite{bronstein2021geometric}. Such a network will contain a sufficiently good solution, yet explores a smaller parameter space while training. For instance, the convolutional neural network (CNN), which is greatly successful for image recognition, is constructed by leveraging the translational symmetry of 2D images. Inspired from this, a number of recent works study the role of symmetry in improving QNN architectures~\cite{larocca_2022, mernyei2022equivariant, skolik2023equivariant, meyer_2023, nguyen_2022, zheng_2023, east2023need}. The results show that a linear map representing the QNN can adapt to a symmetric dataset if its action on the data points commute with the action of the symmetry group (two operators $A$ and $B$ commute if $AB=BA$ or equivalently if $[A,B]\equiv AB-BA=0$). These QNNs are called equivariant quantum neural networks (EQNNs), in analogy to equivariant neural networks in classical geometric learning. EQNNs have less number of trainable parameters and reduced expressibility, but they are expected to show a faster training time and improved generalization compared to a general QNN with high expressibility. Indeed, studies show that some classes of symmetry-respecting QNNs are devoid of the barren plateau problem~\cite{pesah_2021, schatzki2022theoretical, west2023provably}. Moreover, the performance of an EQNN shows improvement over a non-equivariant QNN for pattern recognition and image classification~\cite{meyer_2023, West_2023, chang2023approximately, west2023provably}.

A set of symmetry operations form an abstract group $\mathcal{G}$. A \textit{representation} of a group $\mathcal{G}$ is a mapping $R:\mathcal{G}\rightarrow GL$ where $GL$ is the space of invertible linear operators that preserve the group structure of $\mathcal{G}$. Let us suppose we are given a classical data point with a certain symmetry $\mathcal{G}$. If we encode the data point using an $N$-dimensional quantum state, then the representations of the group elements are $N\times N$ unitary matrices. However more than one unitary representations for the same group is possible. Change of representation is induced by the change in the quantum embedding of the data. In EQNN, the action of the network on the input state must commute with representation $R(g)$ for all group elements $g \in \mathcal{G}$. As different representations have different commutator space, the choice of representation decides which PQC and measurements are to be used in the EQNN. This has been discussed in Ref.~\cite{ragone2023representation} where the authors conceptualize a network in which the representation of the input state is altered at the intermediate layers by applying a linear transformation on the quantum state. Also in Refs.~\cite{West_2023, west2023provably} the authors encode classical images using an altered-basis amplitude embedding in a way so that the resulting group representation simplifies the construction of EQNN. However, whether the resulting EQNN has a similar performance as before the basis-change is not clear. In this work, our objective is to attain a better understanding of how a change in the data embedding affects the performance of EQCNN. We present our theoretical observations about the role of embedding and the resulting representation on the construction of equivariant convolutional and pooling layers in EQCNNs. For our numerical study, we choose standard amplitude embedding of images along with other permuted-basis amplitude embedding, similar to Refs.~\cite{West_2023, chang2023approximately}. We then consider datasets in which the class labels are symmetric with respect to reflection and $180^{\circ}$ rotation. We compare the classification accuracy of a number of EQCNNs with different permuted basis embeddings to a general non-equivariant QCNN for image classification. These results show that the choice of embedding substantially affects the performance of the EQCNN, and can be crucial for obtaining an improvement over non-equivariant QCNNs.

The manuscript is arranged as follows. In Sec. \ref{prelim}, we start with an introduction to QCNN and EQCNN, followed by discussing the role of representation in constructing an EQCNN. In Sec. \ref{our_symmetry} we describe the reflection symmetry group and its different representations used in this work, alongside presenting the unitary ansatze used in building the EQCNNs and the non-equivariant QCNN. We present our findings in Sec. \ref{results} and conclude in Sec. \ref{conclusion}.

\begin{figure}[ht]
    \centering
    \includegraphics[width=0.48\textwidth]{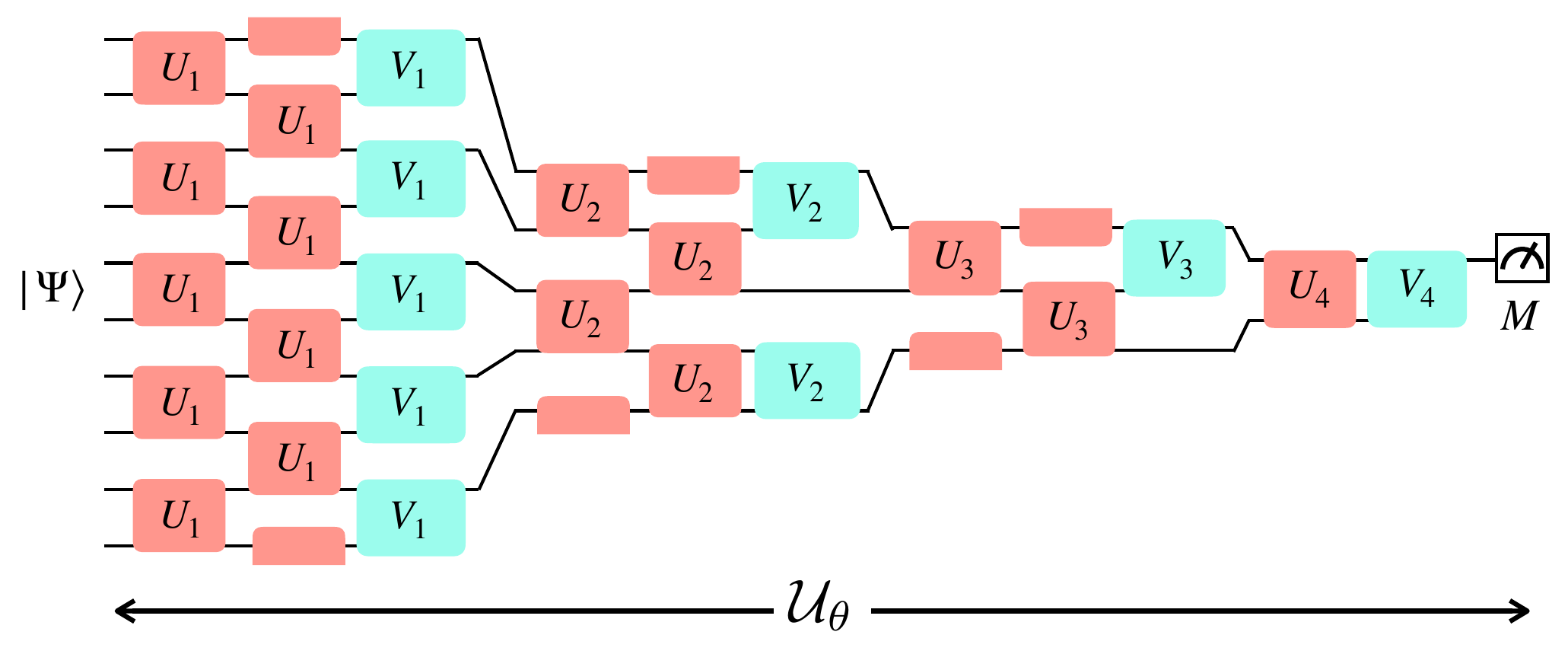}
    \caption{The structure of QCNN for 10 qubits. The orange and the cyan boxes represent respectively the parametrized convolutional and pooling ansatze. The input quantum state is $\vert \psi\rangle$ and the final measurement is $M$.}
    \label{qcnn_layout}
\end{figure}

\section{Equivariant quantum convolutional neural network}
\label{prelim}
\subsection{Quantum Convolutional Neural Network (QCNN)}
Convolutional neural networks (CNNs) are classical ML models extensively used for image classification, speech recognition etc \cite{lecun_2015, SCHMIDHUBER201585}. They consist of a sequence of convolutional and pooling layers followed by a fully-connected layer at the end. The convolution operation can be visualized as a $d\times d$ matrix with trainable weights, known as \textit{kernel}, traversing along the height and width of the input image. At each position of the kernel, the dot product between it and the $d \times d$ block of the input image over which it is placed is calculated to get a feature map of the input image. 
The trainable weights of the kernel are the same within a convolutional layer. It is common to apply a nonlinear activation function after each convolutional layer. In the pooling layer, the dimension is reduced by aggregating over areas of the feature map by taking the average or the maximum value. After application of a number of convolution and pooling layers, the feature maps are flattened to 1D and used as the input layer of a fully-connected neural network that performs the prediction. The cost function is calculated using the output nodes and the network is trained. One important aspect of CNN is that it can recognize a particular feature of an image irrespective of the translation of that feature in the image plane. Thus, CNNs respect the translational symmetry of images.

In Ref.~\cite{cong_2019}, the authors proposed a quantum analogue of CNN. We show this architecture in Fig.~\ref{qcnn_layout}. The architecture is inspired from the multiscale entanglement renormalization ansatz (MERA) representation of quantum many-body states \cite{mera_2008}. Any quantum neural network has three components-- an $n$-qubit quantum register encoding the input quantum state $\vert \Psi\rangle$, a PQC $\mathcal{U}_\theta$ acting on $\vert \Psi\rangle$, and lastly a measurement $M$ on some (or all) of the qubits. In case of QCNN, $\mathcal{U}_\theta$ consists of a series of convolutional and pooling layers. In the $i^{\mathrm{th}}$ convolutional layer, an $m$-qubit ($m<n$) trainable convolutional ansatz $U_{i}(\Theta_{i})$ is applied on all combinations of neighboring $m$ qubits, mimicking the action of a kernel in CNN. Here $\Theta_{i} =\{\theta_{i}^{1}, \theta_{i}^{2}, .., \theta_{i}^{k}\}$ is a set of $k$ trainable parameters which is the same for all $U_{i}(\Theta_{i})$ in the $i^{\mathrm{th}}$ convolutional layer. In the pooling layer, from each pair of neighboring qubits one qubit is measured in a particular basis, and conditioned on the measurement outcomes a set of parametrized rotations are applied on the other qubit. These actions constitute the pooling ansatz $V_{i}(\Phi_{i})$. Similar to convolution layers the parameters $\Phi_{i}=\{\phi_{i}^{1}, \phi_{i}^{2}, .., \phi_{i}^{l}\}$ are shared within a pooling layer. Note that one can also relax the translational invariance condition by using convolutional and pooling ansatze with different parameters within a layer~\cite{liu_2023}. Following the pooling layer, all the measured qubits are traced out, thus reducing the effective dimension of the system. In real quantum devices this is synonymous with ignoring the qubits in the subsequent stages after the measurement, and considering only the dynamics on the remaining qubits. The sequence of a convolutional layer followed by a pooling layer is then repeated until a small fraction of qubits are left. In analogy to the fully-connected part of a CNN, one can then apply a general PQC on the remaining qubits at the end and finally measure them to obtain the prediction. The network is then trained using a suitable loss function and optimization algorithm. Compared to a generic deep quantum neural network, in this case due to progressive qubit reduction, the structure has a shallower depth of $\mathcal{O}(\log n)$, which is preferable for training a QNN. In~\cite{cong_2019} the authors demonstrated the utility of QCNN for topological phase recognition of quantum many-body states as well as optimization of error correcting codes. Later, this architecture of QCNN was as well studied for classical image classification in~\cite{Hur2022}. In this case, there is an additional step in which the classical images are encoded as quantum states and used as the input to the network. In this regard, two widely used embedding methods are amplitude embedding and qubit embedding-- the former encode each classical information value in one of the quantum state amplitudes while the latter employs one qubit to encode one value. For large images amplitude embedding is advantageous as we need logarithmically less qubits compared to number of pixels to encode them. For smaller images, qubit embedding is preferred since the information about each pixel is more easily encoded in the single qubits. In~\cite{Hur2022}, the convolution and pooling layers were applied until only a single qubit remained, on which Pauli-$Z$ measurement is performed. This structure is suitable for binary classification since the expectation value of the measurement operator on a qubit can be associated with the two classes to be distinguished. Overall, good classification accuracy for simple datasets (e.g., MNIST and Fashion MNIST) can be reached. However, one must note that the state-of-the-art QCNN is still too immature to learn the complex features of general complex image datasets. Thus, in general CNNs can achieve higher classification accuracies than QCNNs.

\begin{figure*}[ht]
    \centering
    \includegraphics[width=0.8\textwidth]{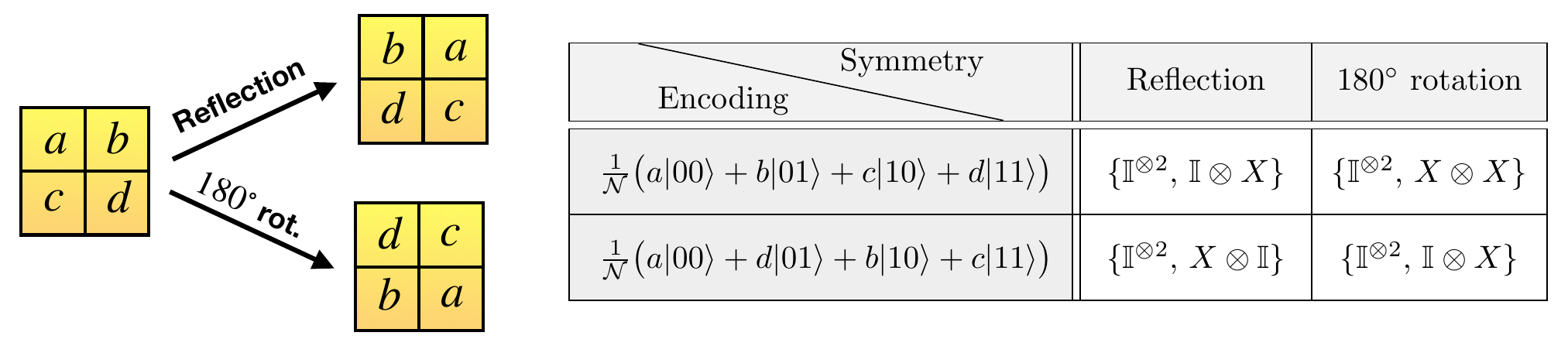}
    \caption{A $2\times 2$ image and its transformations under reflection and rotation by $180^{\circ}$. The table shows group representations of these two symmetries with changing embedding of the image as a quantum state. $\mathcal{N}$ is the quantum state normalization factor.}
    \label{eqcnn_symmetry_table}
\end{figure*}

\begin{figure}[b]
    \centering
    \includegraphics[width=0.25\textwidth]{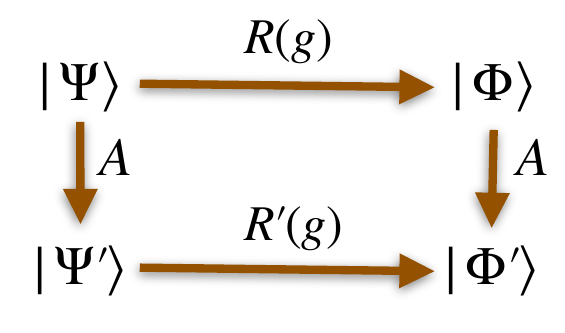}
    \caption{The relation between quantum-encoded image state $\vert \Psi\rangle$, basis-permuting matrix $A$, and the representations $R(g)$ and $R^{\prime}(g)$ before and after the basis permutation, respectively.}
    \label{basis_change_schematc}
\end{figure}

\subsection{Equivariant QCNN with label-symmetry}
The architecture of CNNs respect the translational symmetry of 2D images. However, there can exist additional symmetries in a dataset. Particularly, for classical images one of the most commonly occurring symmetry is \textit{label symmetry} in which the class labels of the images remain unchanged under a set of operations. For example, in MNIST dataset the labels of digits 1, 8 and 0 are reflection invariant. Let us consider a binary-classification task for a dataset $\mathcal{X}$ with images $x_{i}$ and corresponding class labels $y_{i} \in \mathcal{Y} $ where $\mathcal{Y}=\{0, 1\}$. There is a function $f: \mathcal{X}\rightarrow \mathcal{Y}$ that maps the images to their labels. The task of the QCNN is to prepare a quantum circuit $f^{\prime}_{\theta}$ that closely approximates $f$. A set of operations $\mathcal{G}$ form a label symmetry group for $\mathcal{X}$ if for applications of $g\in\mathcal{G}$ to the images, the assigned labels remain unchanged, i.e.,
\begin{equation}
    f(g(x_{i})) = y_{i} = f(x_{i}) \quad \forall x_{i} \in \mathcal{X},\quad\forall g\in \mathcal{G}.
\end{equation} 
Note that the data points themselves may not be invariant under the group action, i.e., $g(x_{i}) \neq x_{i}$ in general. The QCNN $f^{\prime}_{\theta}$ used to classify $\mathcal{X}$ should not predict different labels for inputs related by the symmetry operations. The QCNNs that satisfy this condition are called EQCNNs. Recent works explore EQNN and its improvements over a general non-equivariant QNN for image classification~\cite{West_2023, chang2023approximately}. In particular, we use the result from~\cite{West_2023} which shows that a PQC $\mathcal{U}_{\theta}$ along with a measurement $M$ will construct an EQNN $f^{\prime}_{\theta}$ if the following condition holds,
\begin{equation}
    \big[R(g), \mathcal{U}_{\theta}^{\dagger}M\mathcal{U}_{\theta} \big] = 0 \quad \forall g\in\mathcal{G},
\label{eqv_condition1}
\end{equation}
where $R(g)$ is the unitary representation of $g \in \mathcal{G}$. It was also proved in~\cite{nguyen_2022, ragone2023representation} that a map $\mathcal{U}_{\theta}$ 
and a measurement $M$ will form an EQNN if and only if
\begin{equation}
    [R(g), \mathcal{U}_\theta] = 0 \quad  \mbox{and} \quad [R(g), M] = 0, \quad \forall g
    \label{eqv_condition2}
\end{equation}
Thus if we define the \textit{commutator space} of $R(g)$ to be the space of all operators that commute with $R(g)$, then $\mathcal{U}_{\theta}$ and $M$ must belong to that commutator space. It is easy to see that Eq. (\ref{eqv_condition1}) is satisfied when Eq. (\ref{eqv_condition2}) is true.

In the context of EQCNN, $\mathcal{U}_{\theta}$ can be obtained by assuring that each convolution and pooling layer is equivariant with respect to the representation in the output of the previous layer. Since in our case these layers are composed of two-qubit local variational ansatze, 
it is sufficient to make these ansatze equivariant with respect to the two-qubit local symmetry representations $r(g)$.
In detail, for the $i^{\mathrm{th}}$ layer,
\begin{equation}
    [r(g), U_{i}(\Theta_{i})] = 0 \quad \mathrm{and}\quad [r(g), V_{i}(\Phi_{i})] = 0 \quad \forall r(g).
    \label{eqv_condition3}
\end{equation}
In particular, $U_{i}$ and $V_{i}$ are constituted of parameterized single-qubit and two-qubit rotational gates which can be expressed as $ e^{-i\theta H}$, where $H$ is a Hermitian operator and is called the \emph{generator} of that gate. In this case, Eq. (\ref{eqv_condition3}) is satisfied if 
\begin{equation}
    [r(g), H] = 0
\end{equation}
holds for all the generators of the single and two qubit gates used in the construction of $U_{i}$ and $V_{i}$. Given a symmetry representation $R(g)$, these generators can be found using a variety of methods as discussed in Refs.~\cite{nguyen_2022, meyer_2023, ragone2023representation}.

\subsection{Role of data embedding}

If an image has symmetries, then the representation of that symmetry group depends on the particular way the image is encoded using qubits. In this work, we will consider amplitude embedding (AE) in which $N$ pixel values are encoded as the amplitudes of the basis states of $n=\log N$ qubits. Let us consider two simple label symmetries-- reflection with respect to the vertical axis and rotation by $180^{\circ}$. In standard AE the pixel values of the classical image are encoded row-wise into the amplitudes of the quantum state. Let us consider a small image for which $N=4$ and $n = 2$. For standard AE the representation of reflection group is $\{\mathbb{I}^{\otimes 2}, \mathbb{I} \otimes X\}$ and that for the $180^{\circ}$ rotation group is $\{\mathbb{I}^{\otimes 2}, X \otimes X\}$, where $\mathbb{I}$ is the qubit identity operator and $X$ is the Pauli-$X$ operator. However, if one modifies the order in which the pixels are encoded, in a way presented in the table in Fig.~\ref{eqcnn_symmetry_table}, then the representation for the reflection symmetry group becomes $\{\mathbb{I}^{\otimes 2}, X \otimes \mathbb{I}\}$ and that for the $180^{\circ}$ rotation symmetry group becomes $\{\mathbb{I}^{\otimes 2}, \mathbb{I} \otimes X\}$.

In a more general scenario, let us suppose $A$ is a $2^{n} \times 2^{n}$ matrix that acts on the standard amplitude-encoded input state $\vert \Psi\rangle$ and alters the order in which pixel values are encoded. Thus $A$ has to be a permutation matrix which permutes the coefficients of the canonical basis states of $\mathcal{H}^{n}$, or equivalently permutes the basis states. For input state $\vert \Psi\rangle$ with standard AE the group representation is $R(g)$, while for input state $\vert \Psi^{\prime}\rangle$ with basis-permuted AE, it is $R^{\prime}(g)$. If $\vert \Phi\rangle$ and $\vert \Phi^{\prime}\rangle$ are the states obtained after applying the corresponding symmetry transformations, then the relation between them can be summarized as that in Fig.~\ref{basis_change_schematc}. From this we can write the following--
\begin{eqnarray}
    & R^{\prime}(g)A\vert \Psi\rangle = AR(g)\vert \Psi\rangle \nonumber \\
     \Rightarrow & R^{\prime}(g) = AR(g)A^{-1}.
\end{eqnarray}
In other words, $R(g)$ and $R^{\prime}(g)$ are related by a basis permutation and they are two equivalent representations of the same group. In the same way, the unitary operators in the commutator space of $R(g)$ are related by a basis permutation $A$ to the unitary operators in the commutator space of $R^{\prime}(g)$.

However, the architecture of EQCNNs is a special case of QCNN differing in the following aspects from a general QNN.
\begin{enumerate}[leftmargin=*]
    \item $\mathcal{U}_{\theta}$ is composed of $m$-qubit local unitary ansatze which are equivariant with respect to the locally acting components of the full symmetry representation. Whether these locally equivariant ansatze can realize the full set of globally equivariant ansatze depends on the particular group and its representations~\cite{marvian_2022}.
    \item Due to translational symmetry, in one particular layer of EQCNN all local ansatze must be the same. However, the local symmetry representations may not be same for all groups of $m$ qubits. Let us suppose that in the $i^{\mathrm{th}}$ layer there are $s$ distinct local representations $r_{i}^{j}(g)$ ($j=1, 2,..., s$) and the corresponding sets of equivariant generators are $\sigma_{i}^{j}$. Then the local ansatze in this layer must be generated from the elements which are common to all $\sigma_{i}^{j}$. Thus, the translation symmetry hinders the use of the full set of equivariant generators with respect to the local symmetries and reduces the expressibility of the network. In other words, the EQCNN is capable of taking advantage of only some of the local symmetries that exist in groups of $m$ qubits and not the global or more-than-$m$ qubit symmetries.
    
    \item This dependence on local symmetries becomes more significant after each pooling layer when some of the qubits are traced out. It means the local ansatze are not applied on all the local groups of qubits. 
\end{enumerate}

From the discussion above, it is evident that EQCNN ansatze depend on the local symmetries and the reduced symmetries in the subsequent layers. Thus, a change in representation $R(g)$ is expected to vary the expressibility. We note here again that the changes in representation and the expressibility is accompanied with a basis permutation of the input state itself. Though the set of input states are different for different representations, the correlations between images related by a symmetry operation remain unchanged.

There are a few earlier works where particular basis-permuted embedding was used in order to facilitate construction of the ansatze that satisfy Eq. (\ref{eqv_condition2})~\cite{West_2023, west2023provably}. In this work, our aim is to compare the non-equivariant QCNN with standard AE to EQCNNs with a number of basis-permuted AEs. In order to do this, we first choose a particular $R^{\prime}(g)$ and then we find the corresponding $A$ to be applied to the standard amplitude encoded state $\vert \Psi\rangle$ to get the basis-permuted embedding $\vert \Psi^{\prime}\rangle$. Observing that the matrices that are related by a similarity transformation have the same eigenvalues, we can write
\begin{eqnarray}
   &  M^{-1}R(g)M = N^{-1}R^{\prime}(g)N \nonumber \\
   \Rightarrow & R^{\prime}(g) = NM^{-1}R(g)MN^{-1},
\end{eqnarray}
where $M$ and $N$ are matrices constituted of eigenvectors of respectively $R(g)$ and $R^{\prime}(g)$. Thus, the desired value of $A$ is
\begin{equation}
    A = NM^{-1}.
\end{equation}

In the next section we discuss in detail the architecture of our EQCNN and non-equivariant QCNN for different embeddings.

\begin{figure*}[ht]
    \centering
    \subfigure[]{\includegraphics[width=0.70\textwidth]{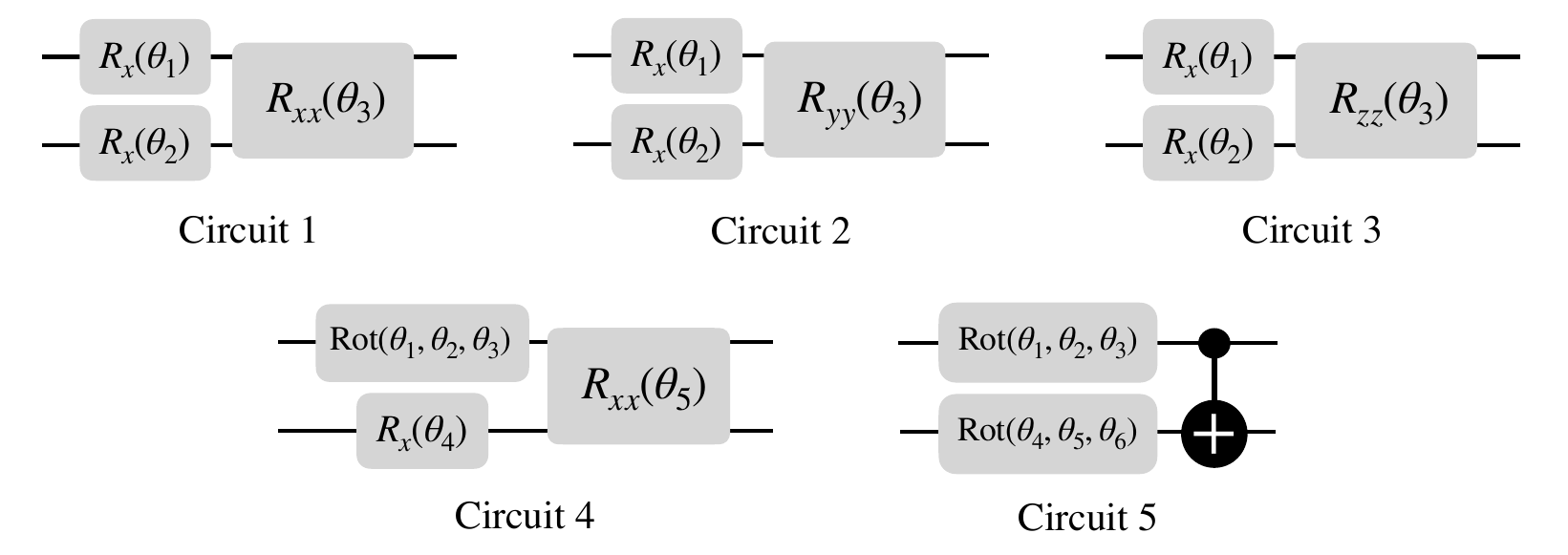}}\vspace{-0.4cm}
    \subfigure[]{\includegraphics[width=0.85\textwidth]{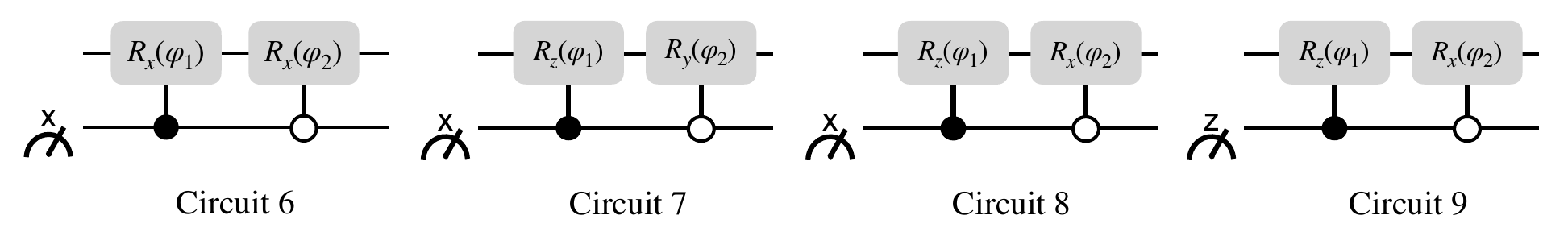}}
    \caption{(a) The convolutional ansatze and (b) the pooling ansatze used in this work. The specific use case of each ansatz is discussed in Sec. \ref{our_symmetry}.}
    \label{ansatze}
\end{figure*}

\begin{table*}[]
\renewcommand{\arraystretch}{1.5}
\begin{tabular}{|C{20mm}|C{13mm}|C{18mm}|cccccccC{8mm}|C{6mm}|}
\hline
\multirow{2}{*}{Dataset}                                                        & \multirow{2}{*}{\# Qubits} & \multirow{2}{*}{Embedding} & \multicolumn{8}{c|}{Ansatz}                                                                                                                                                              & \multirow{2}{*}{M} \\ \cline{4-11}
                                                                                &                            &                            & \multicolumn{1}{C{6mm}|}{U1} & \multicolumn{1}{C{6mm}|}{V1} & \multicolumn{1}{C{6mm}|}{U2} & \multicolumn{1}{C{6mm}|}{V2} & \multicolumn{1}{C{6mm}|}{U3} & \multicolumn{1}{C{6mm}|}{V3} & \multicolumn{1}{C{6mm}|}{U4} & V4 &                    \\ \hline
\multirow{3}{*}{\begin{tabular}[c]{@{}c@{}}Fashion \\ MNIST\end{tabular}}       & \multirow{3}{*}{8}         & AE                         & \multicolumn{1}{c|}{1}  & \multicolumn{1}{c|}{6}  & \multicolumn{1}{c|}{1}  & \multicolumn{1}{c|}{6}  & \multicolumn{1}{c|}{4}  & \multicolumn{1}{c|}{7}  & \multicolumn{1}{c|}{-}  & -  & $\sigma_{Z}$       \\ \cline{3-12} 
                                                                                &                            & AE 1                       & \multicolumn{1}{c|}{1}  & \multicolumn{1}{c|}{6}  & \multicolumn{1}{c|}{2}  & \multicolumn{1}{c|}{6}  & \multicolumn{1}{c|}{3}  & \multicolumn{1}{c|}{6}  & \multicolumn{1}{c|}{-}  & -  & $\sigma_{X}$       \\ \cline{3-12} 
                                                                                &                            & AE 2                       & \multicolumn{1}{c|}{1}  & \multicolumn{1}{c|}{7}  & \multicolumn{1}{c|}{5}  & \multicolumn{1}{c|}{9}  & \multicolumn{1}{c|}{5}  & \multicolumn{1}{c|}{9}  & \multicolumn{1}{c|}{-}  & -  & $\sigma_{Z}$       \\ \hline
\multirow{3}{*}{\begin{tabular}[c]{@{}c@{}}Cifar10/\\ Blood MNIST\end{tabular}} & \multirow{3}{*}{10}        & AE/AE1                     & \multicolumn{1}{c|}{1}  & \multicolumn{1}{c|}{6}  & \multicolumn{1}{c|}{1}  & \multicolumn{1}{c|}{7}  & \multicolumn{1}{c|}{1}  & \multicolumn{1}{c|}{9}  & \multicolumn{1}{c|}{4}  & 8  & $\sigma_{Z}$       \\ \cline{3-12} 
                                                                                &                            & AE 1/AE                    & \multicolumn{1}{c|}{1}  & \multicolumn{1}{c|}{6}  & \multicolumn{1}{c|}{2}  & \multicolumn{1}{c|}{6}  & \multicolumn{1}{c|}{3}  & \multicolumn{1}{c|}{6}  & \multicolumn{1}{c|}{3}  & 6  & $\sigma_{X}$       \\ \cline{3-12} 
                                                                                &                            & AE 2                       & \multicolumn{1}{c|}{1}  & \multicolumn{1}{c|}{7}  & \multicolumn{1}{c|}{5}  & \multicolumn{1}{c|}{9}  & \multicolumn{1}{c|}{5}  & \multicolumn{1}{c|}{9}  & \multicolumn{1}{c|}{5}  & 9  & $\sigma_{Z}$               \\ 
\hline
\end{tabular}
\caption{The datasets, their quantum embeddings, corresponding circuit indices of convolutional and pooling ansatze from Fig. \ref{ansatze} and measurements $M$ used in the construction of EQCNN. Here we use AE, AE 1 and AE 2 to indicate respectively standard AE, basis-permuted AE 1 and basis-permuted AE 2}.
\label{ansatze_table}
\end{table*}

\section{EQCNN with different data embeddings}
\label{our_symmetry}
We consider classification of images whose labels remain invariant under a reflection about the vertical axis and $180^\circ$ rotation. In both cases, the underlying groups have only two group elements-- the identity operation which keeps the image unchanged, and the operation corresponding to reflection or rotation. They form the abstract group $\mathbb{Z}_{2}$. In our experiments, for the reflection-symmetric images we choose to employ the classes 0 (tshirt/top) and 1 (trouser) of Fashion MNIST dataset, and classes 1 (car) and 2 (bird) of Cifar10 dataset. For rotationally symmetric images, we perform classification between classes 1 and 6, as well as between classes 4 and 5 of Blood MNIST dataset. The Fashion MNIST dataset is a collection of 70000 greyscale images of different clothing articles and accessories. Each image has dimension $28 \times 28$ pixels and the dataset has 10 classes in total. The Cifar10 dataset has 60000 RGB images of animals and vehicles divided into 10 classes, each image having dimension $32 \times 32$ pixels. The Blood MNIST dataset is a collection of $32 \times 32$ pixels RGB images of human Blood cells divided into 8 classes. Each class denotes a cell type and contains a few hundred to about a thousand of images.
We downsize the Fashion MNIST images to $16\times 16$ pixels and encode them using 8 qubits. Regarding the Cifar10 and Blood MNIST images, we keep the dimension unchanged but they are transformed from RGB to greyscale and we use 10 qubits for the embedding. In all cases, we use two-qubit convolutional ansatze and assume periodicity in the qubit register i.e. the first and the last qubits are nearest neighbours. 

For the non-equivariant QCNN we use standard amplitude embedding. For constructing $\mathcal{U}_{\theta}$, it is possible to choose from a large number of non-equivariant convolutional and pooling ansatze, which may affect the overall performance of the QCNN. In this work, we build the two-qubit convolutional ansatze by applying a parametrized arbitrary rotation on each qubit and a $CNOT$ gate to entangle them, as shown in circuit 3 in Fig.~\ref{ansatze}, where $\mathrm{Rot}(\theta_{1},\theta_{2},\theta_{3})=R_z(\theta_{3})R_y(\theta_{2})R_z(\theta_{1})$. As the pooling ansatze we use parametrized controlled $z$ and controlled $x$ rotations on the target qubit, when the control qubit is in computational basis states $\vert 1\rangle$ and $\vert 0\rangle$ respectively. This is presented as circuit 9 in Fig.~\ref{ansatze}. In the last step we measure the remaining qubit 1 in $z$ basis.

For the EQCNN, we use standard AE along with two different basis-permuted embeddings, namely basis-permuted \emph{AE 1} and \emph{basis-permuted AE 2}.

\subsubsection{Standard AE}
In standard amplitude embedding of 2D square images using $n$ qubits, the representation of the reflection group is
\begin{equation}
\mathcal{G}_{\mathrm{ref}} \equiv \{\mathbb{I}^{n}, \mathbb{I}_{1} \otimes \mathbb{I}_{2} \otimes ... \otimes \mathbb{I}_{n/2} \otimes X_{(n/2) +1} \otimes... \otimes X_{n}\},
\label{standard_reflection}
\end{equation}
where $\mathbb{I}^n$ is the $n$-qubit identity operator, and 
$\mathbb{I}_{i}$ and $X_{i}$ denote respectively the identity operator and Pauli-$X$ operator applied on $i^{\mathrm{th}}$ qubit. The second element of $\mathcal{G}$ corresponding to reflection operation, leaves the first half qubits unchanged and flips the second half. This is because, in the standard AE of square images, the amplitudes related to a single row span all the bases with fixed values for the first half qubits and all the possible combinations of $0$ and $1$ for the second half qubits. For a rectangular $2^{n_{1}} \times 2^{n_{2}} $ image, the group representation becomes,
\begin{equation}
    \mathcal{G}_{\mathrm{ref}} \equiv \{\mathbb{I}^{n}, \mathbb{I}_{1} \otimes \mathbb{I}_{2} \otimes ... \otimes \mathbb{I}_{n_{1}} \otimes X_{n_{1} +1} \otimes... \otimes X_{n_{1}+n_{2}}\}.
\end{equation}
Similarly, one can check that the representation of $180^{\circ}$ rotation group for standard AE is,
\begin{equation}
\mathcal{G}_{\mathrm{rot}} \equiv \{\mathbb{I}^{n}, X_{1} \otimes X_{2} \otimes ... \otimes X_{n}\}.
\label{standard_rotation}
\end{equation}

\subsubsection{Basis-permuted AE 1}
Basis-permuted AE 1 swaps the representations of reflection and rotation groups in Eq. (\ref{standard_reflection}) and Eq. (\ref{standard_rotation}). We choose a basis-permuting matrix $A$ such that the image is reflected when $X$ is applied on all the qubits. Thus the representation of reflection group becomes,
\begin{equation}
    \mathcal{G}_{\mathrm{ref}} \equiv \{\mathbb{I}^{n}, X_{1} \otimes X_{2} \otimes ... \otimes X_{n}\}.
    \label{all_X}
\end{equation}
For images with rotational symmetry, we apply $A^{-1}$ to permute the basis vectors so that the representation of the $180^{\circ}$ rotation group becomes,
\begin{equation}
\mathcal{G}_{\mathrm{rot}} \equiv \{\mathbb{I}^{n}, \mathbb{I}_{1} \otimes \mathbb{I}_{2} \otimes ... \otimes \mathbb{I}_{n/2} \otimes X_{(n/2) +1} \otimes... \otimes X_{n}\}.
\end{equation}

\subsubsection{Basis-permuted AE 2}
We permute the basis vectors in a way such that the representation of both the reflection and rotation group becomes
\begin{equation}
\mathcal{G}_{\mathrm{ref/rot}} \equiv \{\mathbb{I}^{n}, \mathbb{I}_{1} \otimes X_{2} \otimes \mathbb{I}_{3} \otimes X_{4}\otimes ...\otimes \mathbb{I}_{n-1} \otimes X_{n}\},
\label{alternate_X}
\end{equation}
i.e., the identity operator and Pauli-X operator are applied alternatively on the qubits. The matrix $A$, however, is different for the two groups in this case.\\

\begin{figure*}[ht]
    \centering
    \includegraphics[width=0.98\textwidth]{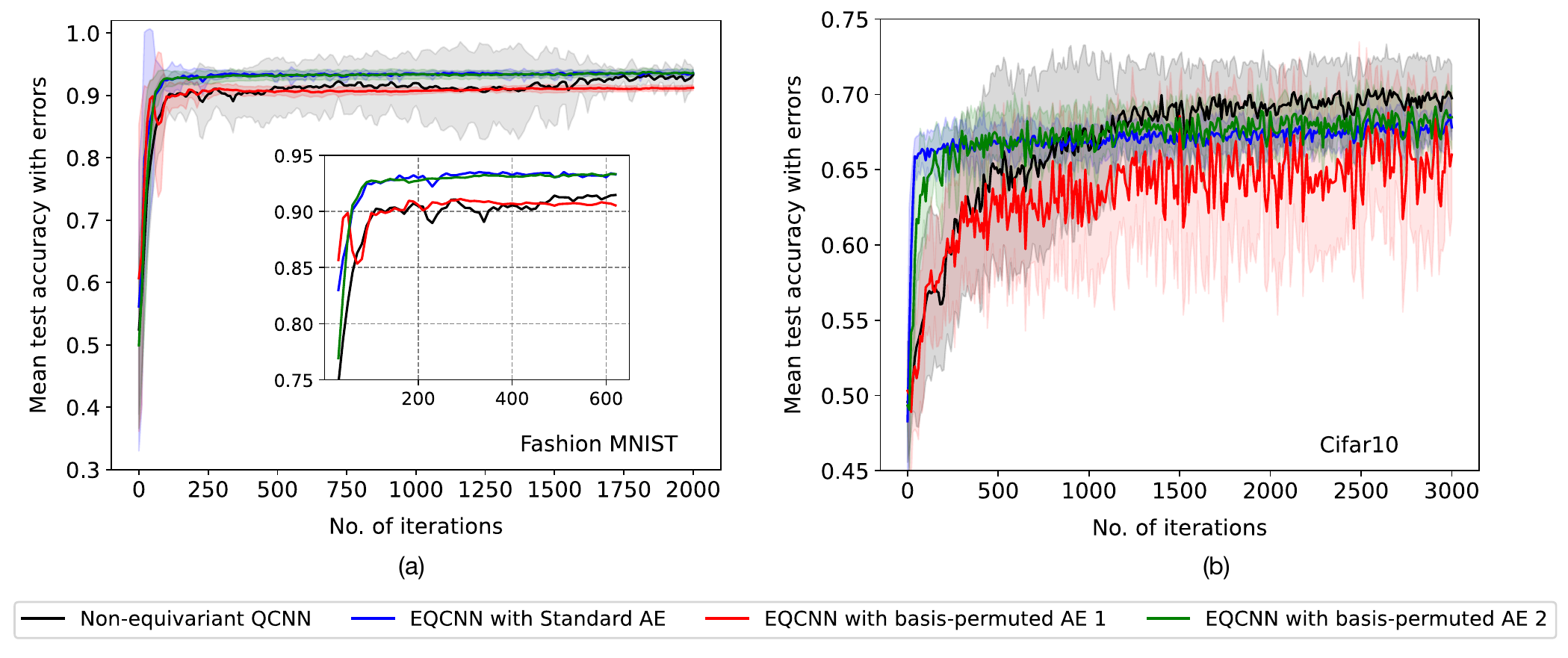}
    \caption{Test set accuracy obtained from equivariant and non-equivariant QCNNs when using maximum possible set of equivariant generators. The vertical axis shows the average accuracy obtained over 10 randomly initialized runs and the standard deviation is indicated with the shaded area. (a) Classification of classes $0$ and $1$ of Fashion MNIST dataset. The inset shows a magnified part of the same with the horizontal axis spanning a few hundreds of iterations. (b) Classification of classes $1$ and $2$ of Cifar10 dataset.}
    \label{reflection_full_gen}
\end{figure*}

Now we describe in detail the construction of the EQCNNs in these three cases. We try to use the largest possible set of equivariant generators in each case for constructing the equivariant convolutional and pooling ansatze, this is summarized in Table \ref{ansatze_table}. However, we also investigate a scenario in which a subset of equivariant generators is used. All the convolutional and pooling ansatze used in this work are presented in Fig. \ref{ansatze}.

Let us consider the standard AE for reflection group for which the representation is given by Eq. (\ref{standard_reflection}). In the first layer of the EQCNN, equivariant convolutional and pooling layers can be built if $U_{1}$ and $V_{1}$ commutes with all the possible two-qubit local group representations within $\mathcal{G}$ which are $\{\mathbb{I} \otimes \mathbb{I}, \mathbb{I} \otimes X, X \otimes \mathbb{I}, X \otimes X\}$. To satisfy the commuting condition, we use circuit 1 in Fig.~\ref{ansatze}(a) for $U_{1}$ and the circuit 6 in Fig.~\ref{ansatze}(b) for $V_{1}$ (note that the control qubit is in the $x$ basis). One can check that these are the only ansatze that commutes with all local representations in this layer. In circuit 6, we have used two $R_{x}$ gates with different parameters to draw an analogy with the other pooling circuits, however one can merge them into a single parametrized $R_{x}$ gate to be applied when the control qubit is either in $\vert 0\rangle$ or $\vert 1\rangle$.
Afterwards, we trace out the even-indexed qubits leaving the reduced representation $\{\mathbb{I}^{5}, \mathbb{I}_{1} \otimes \mathbb{I}_{3} \otimes \mathbb{I}_{5} \otimes X_{7} \otimes X_{9} \}$ for $n=10$ when considering the Cifar10 dataset, and $\{\mathbb{I}^{4}, \mathbb{I}_{1} \otimes \mathbb{I}_{3} \otimes X_{5} \otimes X_{7} \}$ for $n=8$ when considering the Fashion MNIST dataset. The two-qubit local representations for the convolutional ansatze remain unchanged, thus in the second layer we still use circuit 1 for $U_2$. In the second pooling layer, we will trace out qubits 3 and 7 and apply conditional rotations on qubits 1 and 5. The local representations are $\{\mathbb{I}_{1} \otimes \mathbb{I}_{3}, \mathbb{I}_{5} \otimes X_{7}\}$ for $n=10$, and $\{\mathbb{I}_{1} \otimes \mathbb{I}_{3}, X_{5} \otimes X_{7}\}$ for $n=8$. For the former, both circuit 7 and circuit 8 are equivariant generators, however since we can use only one ansatze in a pooling layer, we choose circuit 7 without loss of generality. For  $n=8$ we use circuit 6 as $V_{2}$.  In the third layer, the reduced representations are $\{\mathbb{I}^{3}, \mathbb{I}_{1}\otimes \mathbb{I}_{5} \otimes X_{9}\}$ for $n=10$ and $\{\mathbb{I}^{2}, \mathbb{I}_{1}\otimes X_{5} \}$ for $n=8$. For the former, we use the circuit 1 as $U_3$. For pooling, we trace out qubit 5 by applying conditional rotation on qubit 1, thus $V_{3}$ should commute with $\{\mathbb{I}_{1} \otimes \mathbb{I}_{5}\}$. Here, we are free to use any pooling ansatze, so we choose to use the same one as in the non-equivariant QCNN, i.e. circuit 9. For $n=8$, $U_{3}$ should commute with $\{\mathbb{I}_{1} \otimes X_{5}\}$, thus both circuit 1 and circuit 4 are equivariant ansatze. Since we have already used circuit 1 in previous layers, we use circuit 4 for $U_3$ and the circuit 7 for $V_3$. At this point for $n=8$ we measure the remaining qubit 1 in $z$ basis to get the result. For $n=10$ instead, we still have two qubits left with the reduced representation being $\{\mathbb{I}^{2}, \mathbb{I}_{1} \otimes X_{5}\}$. We apply a further layer with circuit 4 as $U_4$. For the pooling layer, this time we choose circuit 8 as $V_4$. We measure the remaining qubit 1 in $z$ basis.

\begin{figure*}[ht]
    \centering
    \includegraphics[width=0.98\textwidth]{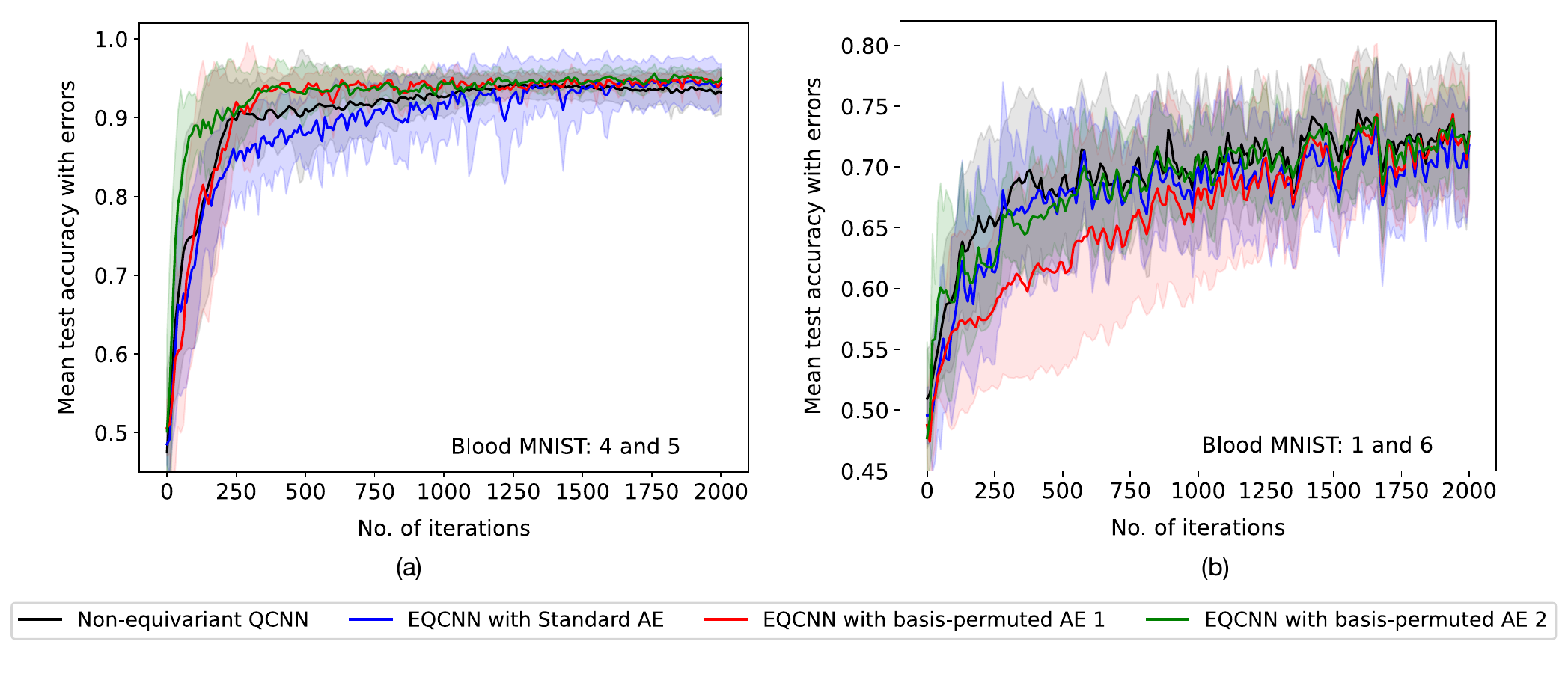}
    \caption{Test set accuracy of Blood MNIST images obtained from equivariant and non-equivariant QCNNs when using maximum possible set of equivariant generators. The vertical axis shows the average accuracy obtained over 10 randomly initialized runs and the standard deviation is indicated with the shaded area.
}
    \label{rotation_full_gen}
\end{figure*}

For $180^{\circ}$ rotation group with standard AE, at every layer the reduced representation is a tensor product of Pauli-X matrices. Thus, $U_{i}$ and $V_{i}$ must commute with $\{\mathbb{I}^{\otimes 2}, X \otimes X\}$ for all layers. The convolutional ansatze that satisfies the latter are circuit 1, circuit 2 and circuit 3 in Fig. \ref{ansatze}(a). Therefore, we use circuit 1 for $U_{1}$, circuit 2 for $U_{2}$, and circuit 3 for $U_{3}$ and $U_{4}$. For pooling ansatze, we are left with the unique choice of circuit 6 for $V_{i}$ at every layer. At the end, to obtain the prediction, we measure the remaining qubit 1 in the $x$ basis. 

For basis-permuted AE 1, we just swap the EQCNN structure for reflection and rotation group discussed above.

For basis-permuted AE 2, in the first layer, $U_{1}$ must commute with $\{\mathbb{I} \otimes X, X \otimes \mathbb{I}\}$ and $V_{1}$ must commute with $\{\mathbb{I} \otimes X\}$, thus we use respectively circuit 1 and circuit 7. In all the subsequent layers, the reduced group representation is simply $\{\mathbb{I}^{\otimes k}\}$, $k$ being the number of qubits in each layer. Since no equivariance constraint is imposed, we use the circuit 5 for $U_i$ and the circuit 9 for $V_i$ for $i>1$. In the end, we measure the remaining qubit in the $z$ basis to get the result.

Note that, one can obtain the same EQCNN for symmetry representations in Eq. (\ref{standard_reflection}) and Eq. (\ref{alternate_X}) by changing the qubits on which the first pooling layer acts. To elaborate, one can choose to trace out qubits $\{6, 7, 8, 9, 10\}$ in the first pooling layer when the representation is that in Eq. (\ref{standard_reflection}). However, we consider a more realistic scenario in which the two-qubit gate connectivity between the qubits is constrained due to the underlying architecture of the quantum hardware. In our case, we implicitly assume that the gate connectivity is exactly the one shown in Fig. \ref{qcnn_layout}. Therefore, our convolutional and pooling layers act on different set of qubits for these two representations.

\begin{figure*}[ht]
    \centering
    \includegraphics[width=0.98\textwidth]{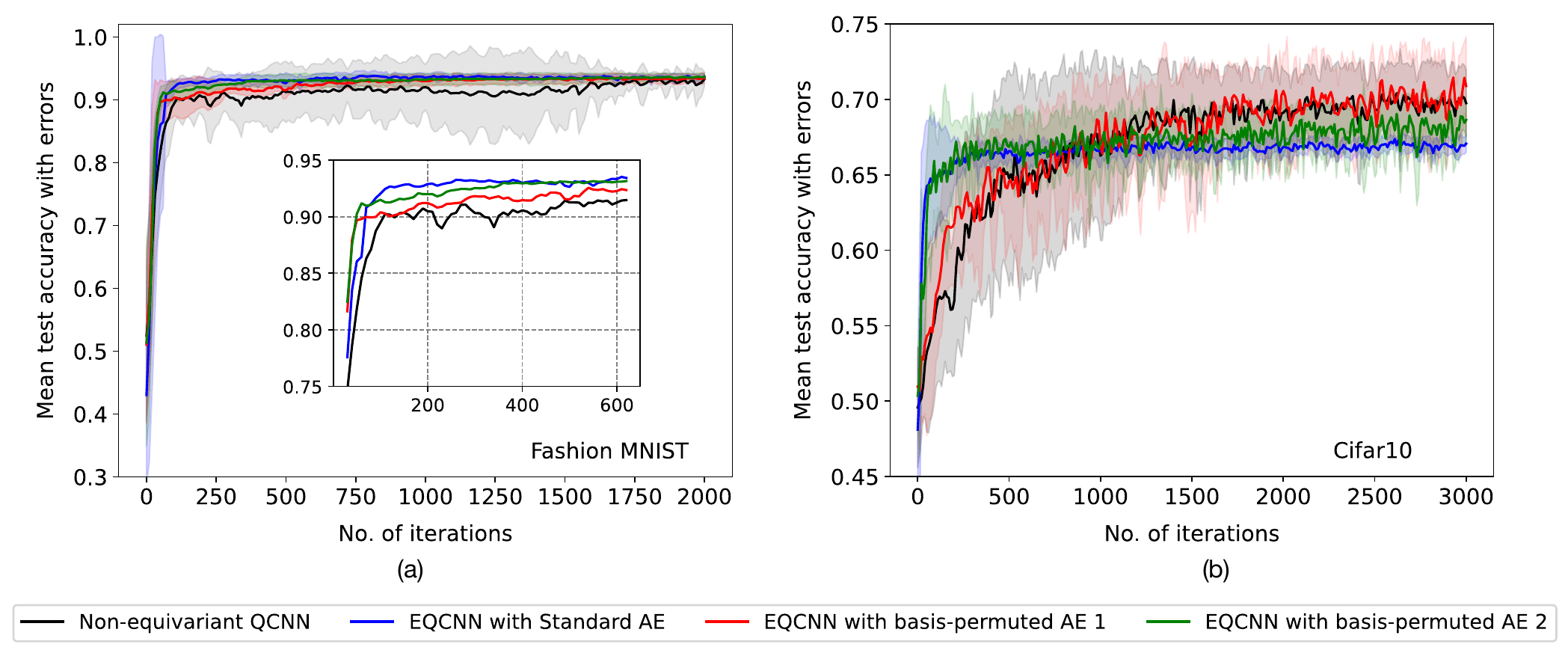}
    \caption{Test set accuracy of (a) Fashion MNIST and (b) Cifar10 images obtained from equivariant and non-equivariant QCNNs when using a subset of equivariant generators. The other details remain same as that in the caption of Fig. \ref{reflection_full_gen}.}
    \label{reflection_sub_gen}
\end{figure*}

\begin{figure*}[ht]
    \centering
    \includegraphics[width=0.98\textwidth]{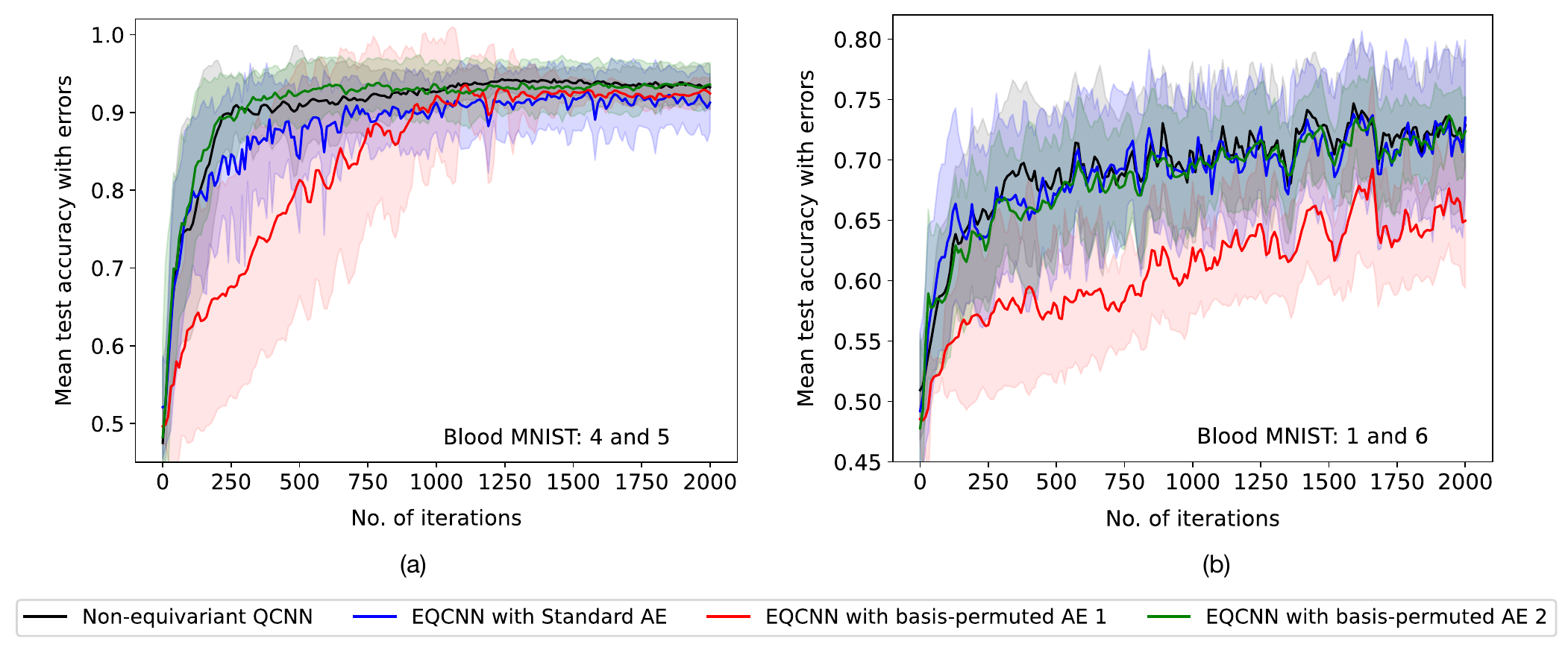}
    \caption{Test set accuracy obtained from equivariant and non-equivariant QCNNs for Blood MNIST dataset when using a subset of equivariant generators. All other details remain same as in the caption of Fig. \ref{rotation_full_gen}.}
    \label{rotation_sub_gen}
\end{figure*}

\section{Results}
\label{results}
We have used Pennylane~\cite{bergholm2022pennylane} quantum simulator to implement the networks. In this work we consider only binary classification. For each dataset we choose a batch size $p$, which is the number of randomly sampled images used for one training iteration. As loss function, we use mean squared error (MSE) which is defined as
\begin{equation}
    C = \frac{1}{p}\sum\limits_{i=1}^{p} (f^{\prime}_{\theta}(x_{i})-y_{i})^{2},
\end{equation}
where $f^{\prime}_{\theta}(x_{i})$ is the expectation value of the measurement operator for the $i^{\mathrm{th}}$ input quantum state in the batch. Since the expectation values of Pauli operators lies in the range $(-1, 1)$, the class labels corresponding to the two classes are mapped to $y_{i}\in \{-1, 1\}$ for all the datasets.
We use Nesterov moment optimizer for training with a learning rate 0.01. We train the network for 2000 or 3000 iterations depending on the dataset. This scheme is run for 10 instances in parallel, for each of which the parameters are randomly initialized. We calculate the average test accuracies and standard deviations over these parallel runs after each 10 iterations.

First, we discuss the results when the maximum possible set of equivariant generators are used. In the following paragraphs, we use `AE 1' and `AE 2' to imply respectively basis-permuted AE 1 and basis-permute AE 2.
For the Fashion MNIST dataset, the training is performed for $2000$ iterations of batches composed of 32 randomly-sampled images. We plot the average test set accuracies and standard deviations for the different equivariant and non-equivariant models with increasing number of iterations in Fig.~\ref{reflection_full_gen}(a). Even though Fashion MNIST is a relatively simple greyscale dataset, the advantage of using an EQCNN over non-equivariant QCNN is clear from the plots. Standard AE and AE 2 achieve a higher accuracy in a much shorter time. The performance of AE 1 lags behind and is comparable to non-equivariant QCNN. All EQCNNs have significantly lower standard deviation compared to the non-equivariant QCNN. 

In Fig.~\ref{reflection_full_gen}(b) we present the test set accuracies for Cifar10 images. In this case, the network is trained for $3000$ iterations with batch size 64. 
We again observe a very fast convergence and higher accuracy for standard AE and AE 2 for lower number of iterations, as well as very low standard deviation. Compared to that, AE 1 shows a slower convergence, lower accuracy and high standard deviation. Overall, non-equivariant QCNN achieves highest accuracy for longer training iterations.

The results for Blood MNIST dataset are presented in Fig. \ref{rotation_full_gen}. We train the network for 2000 iterations with batch size 32 and 64 respectively for classifying between classes $\{4, 5\}$ and classes $\{1, 6\}$. Remember that in this case the representations corresponding to standard AE and AE 1 are swapped. Let us consider the classification of classes 4 and 5. The test accuracies show a similar behaviour to those of reflection-symmetric datasets wherein AE 1 and AE 2 performs better than the others for lower iterations. For longer iterations, all equivariant and non-equivariant QCNNs performs equally well. For classification of classes 1 and 6, in contrast to all other results, AE 1 has significantly lower accuracy for low number of iterations, while all other embeddings performs equally well.

Let us now see how the above trends in accuracy change when using a subset of the equivariant generators. For this, we replace all use-cases of circuit 2 and circuit 3 by circuit 1. Thus, for reflection-symmetric images, the EQCNN corresponding AE 1 is now generated only from $R_{x}$ and $R_{xx}$ gates. The same applies for rotationally-equivariant EQCNN corresponding to standard AE. We also replace all use-cases of circuit 7 by circuit 8, i.e. all pooling layers are generated from $R_{z}$ and $R_{x}$ gates. The resulting behaviour is presented in Fig. \ref{reflection_sub_gen} and Fig. \ref{rotation_sub_gen}. For reflection-symmetric images, this improves the accuracy obtained from AE 1, which either surpasses or matches closely to that obtained from non-equivariant QCNN. For both cases of Blood MNIST dataset, the accuracy of AE 1 significantly decreases.

\begin{figure}[ht!]
    \centering
    \includegraphics[width=0.48\textwidth]{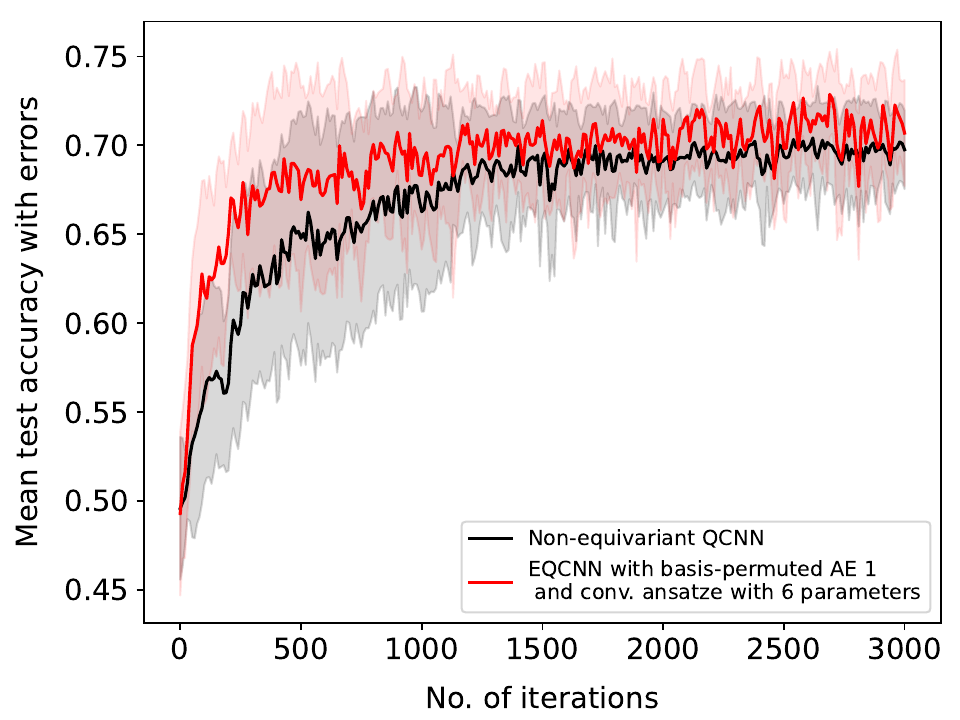}
    \caption{Test set accuracy of Cifar10 for non-equivariant QCNN and EQCNN with 6 trainable parameters in the convolutional ansatze. }
    \label{cifar_comparison_6params}
\end{figure}

We note that the non-equivariant convolutional ansatze has six trainable parameters compared to three trainable parameters in all the equivariant convolutional ansatze. To compare these two QCNNs with an equal number of trainable parameters, we append circuit 2 to circuit 1 to build a new convolutional ansatze with six trainable parameters, and use it in all the layers of EQCNN with AE 1 for classification of Cifar10 dataset. In Fig. \ref{cifar_comparison_6params}, we present the resulting behaviour which shows a higher accuracy for EQCNN compared to the non-equivariant QCNN.

Overall, we observe that the performance of an EQCNN varies depending the classical-to-quantum embedding. In particular, when the group representation is a tensor product of Pauli-X matrices acting on all the qubits, the EQCNN has a lower accuracy. On the other hand, when the representation is a tensor product of Pauli-X matrices acting on half of the qubits and the identity operator acting on the rest half, the EQCNNs have a faster convergence and higher accuracy. For the latter, the pooling choice can make a little difference in the initial training regime, i.e. the EQCNNs show slightly different performance when the pooling layers act on different set of qubits. We note, however, that this behaviour may not show for every dataset, as evident from Fig. \ref{reflection_full_gen}.

\section{Conclusion}
\label{conclusion}
Quantum machine learning, though currently is at its inception, has already been useful in designing novel ML-based algorithms and also has shown some advantages over classical ML. However, there are still a number of open problems when it comes to understanding how to ensure a sufficiently good performance from a QNN. Equivariant QNNs are promising candidates to improve the training and generalization of quantum machine learning algorithms. A typical application of quantum machine learning is image classification which is an ubiquitous task in many daily-life scenarios. For this task, it is possible to construct equivariant QCNN compatible with the label symmetry of images additionally respecting the general translational symmetry of 2D images. In this work, we have explored the connections between the classical-to-quantum embedding of images, the resulting representation of a symmetry group, and the structure of the EQCNN respecting that symmetry. We considered datasets of images characterized by reflection and rotation symmetry and different amplitude embeddings of these images obtained by basis-permutation. Our theoretical observations ascertained that the local representations play a crucial role in deciding the equivariant ansatze to be used and hence the expressibility of the EQCNN. Our numerical results support this by showing a largely varying test set classification accuracy corresponding to different embeddings. It will be interesting to explore if instead of using 2-qubit local ansatze, an $m$-qubit ansatz with $m>2$ can reduce the dependency of the EQCNN on local symmetries. It is also possible to compare amplitude embedding with other kinds of embedding, e.g. qubit embedding and dense qubit embedding \cite{Hur2022}, to investigate their effect on the EQCNN performances. Finally, one can also run the QCNN circuits in the real quantum hardwares, for example by using the corresponding plugins provided by Pennylane. In this case, it will be interesting to see if EQCNNs have better noise robustness compared to non-equivariant QCNNs due to reduced number of parametrized gates used in the former.
\vspace{2em}

\acknowledgments
This work was supported by the European Commission’s Horizon Europe Framework Programme under the Research and Innovation Action GA n. 101070546–MUQUABIS, by the European Union’s Horizon 2020 research and innovation programme under FET-OPEN GA n. 828946–PATHOS, by the European Defence Agency under the project Q-LAMPS Contract No B PRJ- RT-989, and by the MUR Progetti di Ricerca di Rilevante Interesse Nazionale (PRIN) Bando 2022 - project n. 20227HSE83 – ThAI-MIA funded by the European Union - Next Generation EU. S.M. acknowledges financial support from PNRR MUR project PE0000023-NQSTI. S.D. and S.M. thank Paolo Braccia for useful discussions regarding some of the initial ideas.

\bibliography{main}{}

\end{document}